\documentstyle[aps,prl,multicol,epsfig]{revtex}
\catcode`\@=11
\renewcommand{\widetext}{\end{multicols} \global\columnwidth42.5pc}

\newcommand{\beq}{\begin{equation}}
\newcommand{\eeq}{\end{equation}}

\newcommand{\be}{\begin{equation}}
\newcommand{\ee}{\end{equation}}
\newcommand{\bea}{\begin{eqnarray}}
\newcommand{\eea}{\end{eqnarray}}

\newcommand{\ba}{\begin{array}}   %
\newcommand{\ea}{\end{array}}     %
\newcommand{\APPROX}[1]{                
   {{\raisebox{-.3cm}{$\textstyle\simeq$}} \atop {\scriptstyle{#1}}}}
\newcommand{\TO}[1]{                
   {{\raisebox{-.3cm}{$\textstyle\to$}} \atop {\scriptstyle{#1}}}}

\begin{document}
\baselineskip=17pt

\gdef\journal#1, #2, #3, 1#4#5#6{{#1~}{\bf #2}, #3 (1#4#5#6)}
\gdef\ibid#1, #2, 1#3#4#5{{\bf #1} (1#3#4#5) #2}

\title{On the relation between the anyon and Calogero models}

\author{St\'ephane Ouvry }

\date{October 8, 1999}
\address{Laboratoire de Physique Th\'eorique et Mod\`eles 
Statistiques, B\^at. 100, Universit\'e Paris-Sud, 91405 Orsay, France}

\maketitle
\tighten{\begin{abstract}

In order to achieve a dimensional reduction from 
dimension two to one not only in phase space but also in configuration space, the
lowest Landau level (LLL) projection is not sufficient. One has also, in the
LLL, to take the
vanishing magnetic field limit, a procedure which can be given a non ambiguous
meaning by means of  a long distance regulator. As an illustration,
the equivalence of the LLL anyon model in
the  vanishing magnetic field limit to  the Calogero model is established.
A  thermodynamical argument is 
proposed which  supports this claim. Some
general considerations in favor of an intimate connexion between  anyon and
Haldane statistics are also given.

\end{abstract}}
\begin{multicols}{2}
\narrowtext


A few years ago there has been  some discussions aiming  at relating
the two dimensional anyon model \cite{Anyon} to the one dimensional
 Calogero model \cite{Calogero-Sutherland}.
These efforts \cite{Hans} were partly motivated by the fact that both models
 describe identical
particles with statistics continuously interpolating between
Bose-Einstein and Fermi-Dirac statistics.
In the 2d anyon case one speaks of braiding or anyon statistics, in
the 1d Calogero case one speaks of Haldane or exclusion statistics \cite{Haldane}.
The relation between these statistics is
itself an open question since they appear on a quite different footing.
Anyon statistics is microscopically   defined in terms
of a
quantum Hamiltonian whose spectrum, eventhough  not explicitely known,
interpolate continuously between the Bose and Fermi spectra. The interchange
properties  of
the $N$-body eigenstates generalize the Bose and Fermi $\pm$ signature. 
On the other
hand,
Haldane  statistics
is  defined through  a Hilbert space counting argument which
generalizes the Bose and Fermi counting for $N$ identical particles
in $G$ quantum
states of a given energy:  
the number of quantum states available for an additional
particle decreases linearly with
 the number of particles already present. 
 
It happens that the thermodynamics \cite{ES} of
the 1d Calogero model yields 
a mean occupation number which coincides with the one
obtained \cite{Wu} from  Haldane counting for particles with a free 1d
density of states. It is
however not clear 
why particles on a line interacting via $1/x^2$ 
interactions should have an
exclusion like statistics.

On the other hand, if the spectrum of the $N$-anyon model is unknown, 
 a simplification \cite{Hans,Notre,Notrebis}  arises when 
projecting the anyon model onto the lowest Landau level of an external magnetic
field. A complete eigenstate basis, interpolating
between the LLL bosonic and the LLL fermionic  basis, can be found
in the screening regime \cite{Notre} where the flux $\phi$
carried by the anyons  is antiparallel to the 
external magnetic field, i.e. 
when the statistics parameter $\alpha=\phi/\phi_o$, 
which varies from  $\alpha=0$ (Bose) to
$\alpha=\pm 1$ (Fermi), is such that 
$\alpha\in [-1,0]$ if $eB>0$, or
equivalently $\alpha\in [0,1]$ if $eB<0$.
In this situation, 
the LLL anyon thermodynamics
\cite{Notre} turns out to be  similar to those 
of the Calogero model: the mean occupation
number at energy $\omega_c=\vert eB/2m\vert$, i.e. the filling factor in the LLL,
 coincides with the one
obtained from Haldane counting for particles with a infinitely 
degenerate LLL spectrum.
Note however that,
contrary to the Calogero case, the  relation to Haldane statistics 
seems natural here, since
adding an extra anyon implies an additional screening of the external
magnetic field, and thus a Landau degeneracy for the total field (external plus
anyon mean field) decreasing linearly with the number of particles. 

In this Letter, and to come back to the original question of the relation of
the anyon and Calogero models,
one would like to argue that the statement generally advanced in
the literature \cite{Hans}, namely the
equivalence of the Calogero
model and the LLL anyon model, cannot be satisfactory. The LLL anyon model
is  clearly two dimensional: in the thermodynamic limit, its
thermodynamical potential diverges as the surface of the plane \cite{Notre}, with a
$1$-body infinitely degenerate spectrum at energy $\omega_c$. It
 cannot by any means be identical to the 1d Calogero model with a
continuous $1$-body spectrum.
Also, the LLL anyon model is defined in  a screening regime which emphasizes
the importance of the sign of the statistics parameter $\alpha$ with respect
to the orientation of the magnetic field, whereas no track of this
feature can be
found in the Calogero model.

More generally, it is commonly understood that
projecting a  2d system in the LLL makes it
essentially 1d, due to the dimensional reduction of
the $1$-body phase space
from four  to two dimensions.  Numerous applications  have used
this line of reasonning, usually refered to as the Peierles substitution
\cite{Pei}. 
One here argue  that in order to  achieve an actual
dimensional reduction, i.e. not only in phase space but also 
in configuration space, the LLL projection
is not sufficient per se. One  has also, once the system has been
projected onto the
LLL, to take the vanishing magnetic field limit.

It might be objected that taking this limit in the LLL is counter intuitive:
 the LLL projection is
physically justified when the temperature is sufficiently small compared to
the cyclotron gap so that the excited states above the LLL can be ignored.
Thus a strong B field limit is naturally associated 
to the LLL projection, and clearly
such an  interpretation becomes meaningless when the magnetic 
field vanishes.  

Leaving asides its physical interpretation, 
the vanishing $B$ field
limit might not always be formally defined per se, as we will see below.
Still, the procedure proposed here 
can be  given a non ambiguous meaning if some precautions are taken: 
 one has to regularize the system
at long distance, for instance
by means of  a harmonic well of frequency $\omega$ \cite{Reg}, and, only
after i) projecting in the LLL ii)  taking the limit $B\to 0$, can one 
take the
thermodynamic limit $\omega\to 0$.
Under these conditions,   a dimensional reduction of the configuration space
from  dimension two to one is properly achieved.

To illustrate  this line of reasonning -but it should be operative for other
systems as well-, I will indeed show that,
in the vanishing magnetic field limit,  the LLL anyon model
is equivalent to  the Calogero model.
I will conclude  
 by advocating  in favor of an intimate 
connexion  between  anyon and Haldane statistics.

Let me first start by a short reminder on  the anyon model 
 which can be
defined
in the singular gauge by a  free $N$-body Pauli
Hamiltonian ($\hbar=m=1$)
$H_{\rm free}^{u}=-2\sum_{i=1}^N\partial_i\bar\partial_i
\quad,\quad H_{\rm free}^{d}=-2\sum_{i=1}^N\bar \partial_i\partial_i$,
where the index $u,d$ refers here to the spin degree of freedom.
The coupling to an external magnetic field  amounts,
 in the symmetric gauge, to
$\partial\to\partial-eB\bar z/4$ and
$\bar\partial\to\bar\partial+eBz/4$.
The $N$-body
eigenstates $\psi_{\rm free}$ of $H_{\rm free}$ have a non trivial monodromy
encoded in the multivalued phase $ \exp(-i\alpha\sum_{k<
l}\theta_{kl})$
where $\sum_{k< l}\theta_{kl}$ is the
sum of the relative angles between  pairs of particles in the plane. 
 Looking at the monodromy as a (singular) gauge transformation,
one obtains, in the regular gauge,
 a $N$-anyon  Aharonov-Bohm Hamiltonian
acting on monovalued wavefunctions  (bosonic by
 convention)
 with statistics parameter $\alpha=0$ 
for Bose statistics, 
and $\alpha=\pm 1$ for Fermi statistics, with additional  
$\mp\pi\alpha\sum_{i<j} \delta^2(z_i-z_j)$
interactions and $ \mp \sum_i eB/2$  shifts induced by  the
spin up or spin down coupling to the local magnetic field of the
vortices and to the homogeneous magnetic field. 
The short range (contact) interactions
$\mp\pi\alpha\sum_{i<j} \delta^2(z_i-z_j)$ should implement
the exclusion of the diagonal of the
configuration space, and thus have to
be repulsive. So, depending
of the sign of $\alpha$, the spin up Hamiltonian  $(\alpha\in
[-1,0])$
or spin down Hamiltonian $(\alpha\in[0,1])$, is
used.  

Let us concentrate without loss of generality on $\alpha\in[-1,0]$, i.e. on
$H_{\rm free}^{u}$, and, 
in order to compute its thermodynamical properties, let us
  add  a harmonic well  as a long distance regulator.
Thus from now on one considers
\be\label{3} H_{\rm free}=-2\sum_{i=1}^N(\partial_i-{eB\over 4}\bar z_i)
(\bar\partial_i+{eB\over 4}z_i) + 
 \sum_{i=1}^N{\omega^2\over 2}\bar z_i z_i\ee
which describes two different Bose-Fermi interpolations, depending on the
orientation  of
the magnetic field.

To materialize in the eigenstates the short range repulsion  and the long
distance Landau and harmonic exponential damping
one  sets $\psi_{\rm free}=\prod_{k<l}(z_k-z_l)^{-\alpha}\exp(-{1\over
2}\omega_t\sum_{i=1}^N z_i\bar z_i)\psi$
to obtain the Hamiltonian acting on $\psi$
\bea\label{5}
{H} &=& -2  \sum _{i=1}^{N} 
              \left[ \partial_i\bar\partial_i 
                -{\omega_t\pm\omega_c\over 2}\bar z_i\bar \partial_i
		-{\omega_t\mp \omega_c\over 2} z_i \partial_i
		\right]
\nonumber \\
& & +2\alpha\sum_{i<j}\left[{1\over  z_i-
z_j}{(\bar \partial_i-\bar \partial_j})
-{\omega_t\mp\omega_c\over 2}\right]+\sum_{i=1}^N(\omega_t\mp\omega_c)
\eea
where the $\pm$ refers to  the orientation of the
magnetic field (if $eB>0$, $eB/2=\omega_c$,
but if $eB<0$, $eB/2=-\omega_c$) and,
in the presence of the  harmonic well,
$\omega_{\rm t}=\sqrt{\omega_{\rm c}^2+\omega^2}$.

If one puts bluntly $\omega=0$ in (\ref{5}), 
one easily realizes that,  if $eB>0$, i.e. the screening regime,
the Hamiltonian (\ref{5}) acts
trivially on $N$-body eigenstates $\psi$ made of products of 
the 1-body LLL holomorphic eigenstates 
\be \label{6}
({{\omega_c^{\ell_i+1}\over \pi \ell_i!}})^{{1\over 2}}z_i^{\ell_i},\quad
\ell_i\ge 0\ee
of zero energy (remember the
  LLL  spectrum has been  shifted
  downward by the spin induced  $-\omega_c$, thus the LLL has zero energy).
This is the LLL anyon model with an infinitely 
degenerate $N$-body spectrum $E_N=0$.
Note on the other hand that if $eB<0$, (\ref{5}) would not have a simple form
when acting on
 products of  1-body LLL anti-holomorphic eigenstates. 

To proceed further in the  case of interest, i.e. the screening regime,
  one has to recognize \cite{Notre} that the virtue of the harmonic confinement
  $\omega\ne 0$
is precisely to lift the
degeneracy with respect to the $\ell_i$'s and
 dress the $N$-body spectrum with an explicit  $\alpha$ dependence.
In a harmonic well, the $1$-body LLL eigenstates (\ref{6}) become the
$1$-body harmonic LLL eigenstates
\be \label{88}
({{\omega_t^{\ell_i+1}\over \pi \ell_i!}})^{{1\over 2}}z_i^{\ell_i},\quad
\ell_i\ge 0\ee
with  now a non degenerate spectrum
\be\label{10}(\omega_t-\omega_c)(\ell_i+1), \quad \ell_i\ge 0
\ee
Up to a $\omega_t$ dependant normalization,
  the LLL anyonic eigenstates in a harmonic well are symmetrized  
   products of the
$1$-body harmonic LLL eigenstates (\ref{88})  $(0\le\ell_1\le\ldots \le \ell_N )$
    \be\label{8}
 \psi_{\rm free}=\prod_{i<j} (z_i-z_j)^{-\alpha}\prod_{i=1}^N z_i^{\ell_i}
 \exp(-{1\over 2}\omega_t\sum_{i=1}^N z_i \bar z_i)
\ee
Acting on this basis, the Hamiltonian (\ref{5}) rewrites
\be\label{11}
{H}_{LLL} = (\omega_t-\omega_c) \left[\sum_{i=1}^N
z_i \partial_i -\alpha N(N-1)/2+N\right]
\ee
with a harmonic LLL $N$-body spectrum
\be\label{9} E_N=(\omega_t-\omega_c)
\left[
\sum_{i=1}^N
\ell_i
-{1\over2}N(N-1)\alpha+N\right]\ee
which is the sum of the $1$-body harmonic LLL spectra
shifted by the $2$-body statistics term
$-{1\over2}N(N-1)\alpha(\omega_{\rm t}
-\omega_{\rm c})$.
The spectrum
and the eigenstates (\ref{8},\ref{9}) interpolate from the harmonic
LLL bosonic to the
harmonic LLL fermionic basis when $\alpha: 0\to -1$
and lead, in the thermodynamic limit $\omega\to 0$, to Haldane
exclusion statistics with parameter $g=-\alpha$
for a degenerate $1$-body LLL spectrum \cite{Notre}.

At this point, being in the LLL and  a harmonic well, let us take, as
advocated above, the $B\to
0$ limit, i.e. one considers (\ref{5},\ref{88}-\ref{9})
with $B=0$ and $\omega_t= \omega$.
In this limit,  (\ref{5}) becomes of course
 the $N$-anyon Hamiltonian in a harmonic well, here for
$\alpha\in [-1,0]$, bearing in mind that  in the absence of an external
magnetic field the Bose-Fermi interpolations $\alpha: 0\to 1$ and 
$\alpha:0\to -1$ are
 equivalent. The harmonic LLL basis (\ref{88}) become
\be\label{13} ({{\omega^{\ell_i+1}\over \pi \ell_i!}})^{{1\over 2}}
z_i^{\ell_i},\quad  \ell_i\ge 0\ee
with $1$-body spectrum
\be\label{14}\omega (\ell_i+1),\quad  \ell_i\ge 0
\ee
that is one picks up on each 2d harmonic energy level
$(j+1)\omega$, $j\ge 0$,  with degeneracy $j+1$,
the state of maximal angular momentum $\ell=j$, and consequently zero radial
quantum number, yielding (\ref{14}) which happens to coincide with a 1d harmonic 
spectrum. The $N$-body eigenstates become, up to a $\omega$-dependant
normalization 
    \be\label{15}
 \psi_{\rm free}=\prod_{i<j} (z_i-z_j)^{-\alpha}\prod_{i=1}^N z_i^{\ell_i}
 \exp(-{1\over 2}\omega\sum_{i=1}^N z_i \bar z_i)\ee
with the  projected Hamiltonian
\be\label{17}
{H}_{\omega} = 		\omega \left[\sum_{i=1}^N
z_i \partial_i -\alpha N(N-1)/2+N\right]
\ee
and $N$-anyon spectrum
\be\label{16} E_N=\omega\left[
(\sum_{i=1}^N
\ell_i
-{1\over2}N(N-1)\alpha+N\right]\ee
Now, looking at (\ref{16}), one recognizes (up to a global
one-body shift $\omega/2$) the $N$-body 1d Calogero spectrum in a harmonic
well. Moreover, one can show that the Hamiltonian (\ref{17}) is equivalent to the 1d
harmonic Calogero Hamiltonian by formally following
the steps of \cite{Hans}: the algebra of
annihilation-creation operators of the Calogero model in a harmonic well 
with  coupling
constant-exclusion parameter $g=-\alpha\in [0,1]$ can be realized in a
2d holomorphic representation which precisely yields the Hamiltonian
(\ref{17}).

What has been obtained here by taking the $B\to 0$ limit in the LLL
is a projection from a 2d model to a 1d model,
contrary to what a LLL projection alone can achieve.
Note that, looking at the eigenstates (\ref{13}-\ref{15}) and 
forgetting that they were
 obtained from the $B\to 0$ limit in the LLL,  
one has here a ``harmonic'' projection which maps the anyon model on the
Calogero model, without any reference to a magnetic field.

The equivalence obtained at the Hamiltonian and spectrum levels between 
the $B\to 0$ LLL anyon  model and the Calogero model can be
also seen in a thermodynamical approach.
Since, in the thermodynamic limit $\omega\to 0$,  the
2d
harmonic well regulator $1$-body partition function  should become the  
2d free partition function
$Z_o^{d=2}=V/(2\pi\beta)$, one infers that  \cite{Reg}
\be\label{18}
Z= {e^{-\beta\omega}\over (2\sinh {\beta\omega\over 2})^2}\APPROX{\omega\to 0}
{1\over (\beta\omega)^2}\to {V\over 2\pi\beta}\ee
where $V$ stands for the infinite area
of the plane. It follows that 
 the harmonic LLL  $1$-body
partition function corresponding to (\ref{10}) 
becomes in the thermodynamic
limit $\omega\to 0$ (ignoring the global shift)
\be\label{20} {e^{-\beta\omega_t}\over
1-e^{-\beta(\omega_t-\omega_c)}}\APPROX{\omega\to 0} {e^{-\beta\omega_c}\over 
\beta(\omega_t-\omega_c)}\to
e^{-\beta\omega_c} \left|{eB\over 2\pi}\right| V\ee
i.e. of course  the 2d LLL partition function $Z_{LLL}=e^{-\beta\omega_c}
\left|{eB/ (2\pi)}\right|V $  with infinite Landau degeneracy
$BV/\phi_o$. On the contrary,
the ``projected'' 1-body partition function 
obtained by restricting the 2d harmonic spectrum to (\ref{14})
\be\label{19} {e^{-{\beta\omega\over 2}}\over 2\sinh {\beta\omega\over 2}}
\APPROX{\omega\to 0} {1\over \beta\omega}\to \sqrt{V\over 2\pi\beta}\ee
behaves as the  free 1d partition function $Z_o^{d=1}={L/\sqrt{ 2\pi\beta}}$
provided that  $\sqrt V$ is interpreted as the
infinite length $L$ of the
resulting 1d system.

Therefore, in the thermodynamic
limit $\omega\to 0$, the harmonic LLL partition function obtained
with the spectrum (\ref{10})
leads, when $B\ne 0$, to $Z_{LLL}$, and  when $B=0$,
to $Z_o^{d=1}$ meaning that 
the following identity for the partition functions
holds 
\be\label{21} Z_{LLL}=e^{-\beta\omega_c} \left|{eB\over 2\pi}\right| V
\TO{B\to 0}Z_o^{d=1}=\sqrt{{V\over 2\pi\beta}}\ee
and accordingly for the density of states
\be\label{22} \rho_{LLL}=\left|{eB\over 2\pi}\right|
V\delta(E-\omega_c)\TO{B\to 0}\rho^{d=1}_o={\sqrt{V}\over\pi\sqrt{2E}}
\ee
where $\rho_{LLL}$ and $\rho^{d=1}_o$
stand respectively for the LLL and the free 1d density of states. 
Clearly, setting  bluntly
$B=0$ in $Z_{LLL}$ has no meaning whatsoever.
Still, (\ref{21},\ref{22}) have been given a non ambiguous meaning
through the long distance harmonic regularization.
Accordingly, when $B=0$, the LLL eigenstates basis (\ref{6}) is mapped on 
the free 1d plane wave basis. 

The same reasonning equally applies to
the thermodynamics of the LLL anyon model,
with in  dimension  two a thermodynamic limit
prescription  \cite{Notre,Reg} at order $n$ in
the cluster expansion
$1/( \beta \omega)^2 \to nV/(2\pi\beta)$
that generalizes (\ref{18}).
The cluster coefficients of the harmonic LLL anyon model, corresponding to
the $N$-body spectrum (\ref{9}),
rewrites in the small $\omega$ limit \cite{Notre} 
\be\label{27} b_n^{LLL}={1\over \beta(\omega_t-\omega_c)}
{e^{-n\beta(\omega_t-\omega_c)}\over
n^2}\prod_{k=1}^{n-1}{k+n\alpha\over k}\ee
Again, one can take the thermodynamic limit,
either with $B\ne 0$, or $B=0$, to arrive respectively at 
the thermodynamical potential of the 2d LLL anyon model and of the 1d Calogero model
\be\label{29} \ln Z=\int dE \rho(E)\ln \Xi(E)\ee
where $\rho$ stand respectively for the 2d LLL density of states
$\rho_{LLL}$
and the 1d free density of states $\rho_o^{d=1}$, the latter being 
as advocated above, the
vanishing magnetic field limit of the former. 
$\Xi$, which may be interpreted as the grand partition function at energy $E$,
satisfies the transcendental equation \cite{ES,Wu,Notre}
\be\label{30}\Xi-x\Xi^{1+\alpha}=1\ee
 $x$ being the Gibbs factor
at energy $E$. 
Equation (\ref{30}), considered as the cornerstone of Haldane-exclusion
statistics, has just been shown to come directly from the thermodynamics
of the 2d LLL anyon model projected either on the harmonic LLL basis, 
or on the ``harmonic'' basis, the latter projection being also understood as
the vanishing magnetic field limit of the former.

To conclude, let us mention that, in the paradygmatic 
LLL anyon case,  
the basic structure of the cluster coefficient (\ref{27}) is, in the thermodynamic limit, 
\be\label{31} b_n=G
{e^{-n\beta E_o}\over
n}\prod_{k=1}^{n-1}{k-ng\over k}\ee
where the infinite Landau degeneracy has been denoted ed by $G$,
the number of quantum states at a given energy $E_o$, and the anyonic
parameter $-\alpha$ has been replaced by the more familiar Haldane exclusion
parameter $g$.
(\ref{31}) as well of the resulting
$N$-body partition function (for a now finite $G$)
\be\label{32} Z_N=G
e^{-N\beta E_o}
\prod_{k=2}^{N}{k+G-1 -N g\over k}\ee
are a direct consequence of the transcendental equation (\ref{30}), expanded
in powers of $x$, with the
grand-partition function given from (\ref{29}) by $Z=\Xi^G$.
Note that 

i) the counting deduced from  (\ref{32})  differs \cite{MS} from the standard 
Haldane counting

ii) if one insists on interpreting $\Xi$  as the grand
partition for a system of exclusion particles in a single state of energy
$E_o$, i.e. $G=1$, which is in principle not allowed since  from
thermodynamics  on average $<N>/G <1/g$, then one has to face in (\ref{32}) the
so-called problem of negative probabilities \cite{Na,MS}, which would be
avoided if one sticks, as it should,  to $G>Ng$.

So one has on the one hand the 2d anyon model
microscopically defined from first principles, 
on the other hand  Haldane statistics
coming from  counting considerations in the Hilbert space.
The thermodynamics of Haldane statistics leads to
the basic equations (\ref{30},\ref{31},\ref{32}),
which, on the other hand, follow directly from  the LLL-anyon model, a particular
limit of which yields, as already said,  yet another microscopic 
example of Haldane statistics, the
Calogero model.  
It is thus tempting to propose  quite generally 
that Haldane statistics might in fact be a 
particular limit of anyon statistics, each time it can be realized in terms
of a microscopic Hamiltonian. 

 I would like to thank C.~Furtlehner, S.~Isakov  and C.~Texier 
for useful discussions.

\end{multicols}

\vspace{-0.5cm}
\begin{references}
\vspace{-1.5cm}

\bibitem{Anyon}  
J. M. Leinaas and J. Myrheim, Nuovo Cimento {\bf B37}, 1 (1977); G. A.
Goldin, R. Menikoff and D. H. Sharp, J. Math. Phys. {\bf 22}, 1664 (1981);
F. Wilczek,  Phys. Rev. Lett. {\bf 48}, 1144 (1982); {\bf 49}, 957 (1982)


\bibitem{Calogero-Sutherland} 
F. Calogero, J. Math. Phys. {\bf 10}, 2191 (1969); {\bf 12}, 419 (1971);
B. Sutherland,  Phys. Rev. {\bf A4}, 2019 (1971); {\bf A5}, 1372 (1972)

\bibitem{Hans} T.H. Hansson, J.M. Leinnaas and J. Myrheim, Nucl. Phys. 
{\bf B384}, 559 (1992);
L. Brink, T.H. Hansson, S. Konstein and M.A. Vasiliev, 
Nucl. Phys. {\bf B401}, 591 (1993) and references therein 


\bibitem{Haldane} 
F. D. M. Haldane, Phys. Rev. Lett. {\bf 67}, 937 (1991)

\bibitem{ES} S. B. Isakov, \journal Int. J. Mod. Phys. A, 9, 2563, 1994 


\bibitem{Wu} S. B. Isakov,
\journal Mod. Phys. Lett. B, 8, 319, 1994; Y.-S. Wu, \journal Phys. Rev.
Lett., 73, 922, 1994

\bibitem{Notre} 
A. Dasni\`eres de Veigy and S. Ouvry, Phys. Rev. Lett. {\bf 72}, 600 (1994); Mod. Phys. Lett.
{\bf B9}, 271
(1995); Phys.
Rev. Lett. {\bf 75}, 352 (1995)

\bibitem{Notrebis} see also
M. D. Jonhson and G. S. Canright, Phys. Rev. {\bf B9}, 2947 (1994); J. Phys.
{\bf A 27}, 3579 (1994)

\bibitem{Pei}  R. Peirles, Z. Phys. {\bf 80}, 763 (1933); for recent
developments see G. Dunne and R.
Jackiw, hep-th/9204057

\bibitem{Reg}  
A. Comtet, Y. Georgelin and S. Ouvry, J. Phys. {\bf A}: Math. Gen. {\bf 22}, 
3917 (1989);
K. Olaussen,
cond-mat/9207005.

\bibitem{MS} A. P. Polychronakos, Phys. Lett. {\bf B365}, 202 (1996);
  M.V.N. Murthy and R. Shankar, ``Exclusion Statistics: A
resolution of the problem of negative weights'',
IMSc/99/01/02 report where  a  relation of (\ref{30}) 
to Ramanujan  equations is discussed; M.C. Berg\`ere, ``Fractional
Statistics'', Saclay report (1999)


\bibitem{Na} C. Nayak and F. Wilczek, Phys. Rev. Lett. {\bf 73}, 2740  (1994)




\end{references}
\end{document}